\documentclass[%
 reprint,
 superscriptaddress,
 amsmath,amssymb,
 aps,
 pra,
]{revtex4-2}

\usepackage[utf8]{inputenc}
\usepackage[T1]{fontenc}
\usepackage{textcomp}       % Provides \texteuro
\usepackage{eurosym}        % Provides \euro
\usepackage{graphicx}% Include figure files
\usepackage{dcolumn}% Align table columns on decimal point
\usepackage{bm}% bold math
\usepackage{upgreek}
\usepackage{xcolor}
\usepackage{newunicodechar}
\usepackage{ulem} % Provides \uline for underlining
\newunicodechar{ }{\,}
\usepackage{hyperref}
\hypersetup{hidelinks}
\begin{document}
\title{Engineered mode coupling in high-Q microresonators enables deterministic low-repetition-rate soliton microcombs}

\author{Yi Zheng}
\thanks{yizhen@dtu.dk}
\affiliation{DTU Electro, Department of Electrical and Photonic Engineering, Technical University of Denmark, 2800 Kongens Lyngby, Denmark}

\author{Yang Liu}
\affiliation{DTU Electro, Department of Electrical and Photonic Engineering, Technical University of Denmark, 2800 Kongens Lyngby, Denmark}

\author{Haoyang Tan}
\affiliation{DTU Electro, Department of Electrical and Photonic Engineering, Technical University of Denmark, 2800 Kongens Lyngby, Denmark}

\author{Yanjing Zhao}
\affiliation{DTU Electro, Department of Electrical and Photonic Engineering, Technical University of Denmark, 2800 Kongens Lyngby, Denmark}

\author{Andreas Jacobsen}
\affiliation{DTU Electro, Department of Electrical and Photonic Engineering, Technical University of Denmark, 2800 Kongens Lyngby, Denmark}

\author{Kresten Yvind}
\affiliation{DTU Electro, Department of Electrical and Photonic Engineering, Technical University of Denmark, 2800 Kongens Lyngby, Denmark}

\author{Minhao Pu}
\thanks{mipu@dtu.dk}
\affiliation{DTU Electro, Department of Electrical and Photonic Engineering, Technical University of Denmark, 2800 Kongens Lyngby, Denmark}

\date{\today}% It is always \today, today,

%\linenumbers

\begin{abstract}
Soliton optical frequency combs have become key enablers for a wide range of applications, including telecommunications, optical atomic clocks, ultrafast distance measurements, dual-comb spectroscopy, and astrophysical spectrometer calibration, many of which benefit from low repetition rates. However, achieving such low-repetition-rate soliton microcombs is nontrivial as long cavities require substantially higher pump power, which induces stronger thermal effects that, in turn, exacerbate thermal instability and complicate access to stable soliton states. The dual-mode pumping scheme, in which a continuous-wave pump couples to both the comb-generating mode and an auxiliary mode, has proven simple and effective for mitigating thermal instability and enabling thermally accessible soliton generation. Yet, in long-cavity devices, the standard bus-to-resonator coupling conditions for these two modes diverge substantially, resulting in insufficient pump coupling to the auxiliary mode, which makes dual-mode pumping particularly challenging for low-repetition-rate microcombs. In this work, we overcome this limitation by coupling the pump to the auxiliary mode via intermodal coupling, which can be introduced in racetrack microresonators and engineered by tailoring the cavity bend design. We validate this approach in a high-Q (>$10^7$) silicon nitride microresonator and demonstrate thermally accessible, deterministic single-soliton generation at a repetition rate of 33 GHz. This work provides a simple and robust pathway for generating low-repetition-rate soliton microcombs.
\end{abstract}

\maketitle

%\tableofcontents

\section{\label{sec:intro}Introduction}

Chip-scale optical frequency combs are transforming classical and quantum applications by offering high-performance coherent light sources with reduced size, weight, and power (SWaP) \cite{Kippenberg2018DissipativeMicroresonators, Gaeta2019Photonic-chip-basedCombs, Wilmart2019AApplications, Liu2021High-yieldCircuits, Ye2023FoundryCircuits, Ji2024EfficientCircuits}. Among these, low-repetition-rate microcombs (typically <50 GHz) are particularly critical for Dense Wavelength Division Multiplexing (DWDM) systems, where they provide the densely spaced, phase-coherent carriers required to maximize data capacity in ultra-long-haul optical transmissions \cite{Marin-Palomo2017Microresonator-basedCommunications, Hu2018Single-sourceTransmissionb, Jrgensen2022Petabit-per-secondSource}. Beyond optical communications, low-repetition-rate microcombs have also attracted considerable interest in dual-comb spectroscopy \cite{Suh2016MicroresonatorSpectroscopy}, astrophysical spectrometer calibration \cite{Obrzud2019}, low-noise microwave generation \cite{Liang2015HighOscillator, Lucas2020Ultralow-noiseOscillator, Liu2020PhotonicMicrocombs}, and narrow-linewidth laser synthesis \cite{Jin2021Hertz-linewidthMicroresonators}. Despite the development of various high-Q integrated platforms, including Si \cite{Griffith2015Silicon-chipGeneration}, $\mathrm{Si_3N_4}$ \cite{Levy2010, Guo2017UniversalMicroresonators}, Hydex \cite{Razzari2010}, diamond \cite{Hausmann2014DiamondPhotonics}, AlN \cite{Jung2014GreenResonator, Liu2023AluminumOptics}, AlGaAs \cite{Pu2016a, Chang2020Ultra-efficientMicroresonators}, $\mathrm{LiNbO_3}$ \cite{Zhang2017MonolithicResonator, He2019Self-startingMicrocomb}, GaP \cite{Wilson2020IntegratedPhotonics}, GaN \cite{Stassen2019High-confinementPhotonics, Zheng2022IntegratedPhotonics}, and SiC \cite{Lukin20204H-silicon-carbide-on-insulatorPhotonics, Zheng20194H-SiCPhotonics, Cai2022Octave-spanningPlatform}, a fundamental physics-to-engineering bottleneck arises from thermal instability. Stable soliton operation requires all comb modes to reside on the red-detuned side of their resonances, which is an intrinsically thermally unstable regime. 

To date, overcoming thermal instability has relied heavily on active intervention. Techniques such as fast frequency sweeping (at speeds up to 100 GHz/$\mu$s) \cite{Guo2017UniversalMicroresonators, Stone2018ThermalCombs}, power kicking \cite{Yi2016}, or auxiliary laser heating have been successfully employed to access the soliton state \cite{Zhou2019SolitonMicrocavities, Zhao2021SolitonLaser, Zhao2025ThermalGeneration}. However, these methods introduce significant system complexity, requiring high-speed electronics or secondary laser sources that complicate integration. Passive mitigation through dispersion and cavity-length design has been explored, but the method is highly dependent on the free spectral range (FSR) \cite{Wu2023AlGaAsTemperature, Jacobsen2025High-powerGeneration}. A more elegant approach is dual-mode pumping \cite{Li2017StablyRegime, Weng2021DirectlyMicroresonator, Weng2022Dual-modeSolitons}, which uses an auxiliary spatial mode to provide thermal compensation. Nonetheless, this approach has thus far been limited to high-repetition-rate (THz-scale) devices \cite{Li2017StablyRegime, Weng2021DirectlyMicroresonator, Weng2022Dual-modeSolitons}, as it requires strong pump coupling to both the comb-generating and auxiliary modes. Extending dual-mode pumping to low-repetition-rate soliton microcombs is considerably more challenging.

\begin{figure*}
  \centering
  \includegraphics[width=11.5cm]{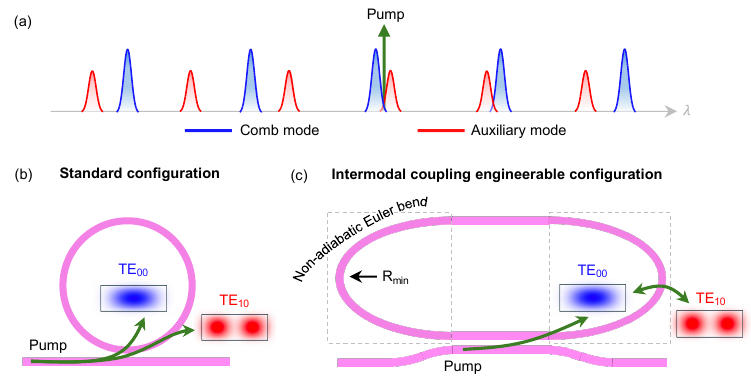}
\caption{\label{fig:principle}\textbf{Engineered intermodal coupling enabling dual-mode pumping for low-repetition-rate soliton microcomb generation.} (a) Schematic of a microresonator spectrum showing two spatial modes, where the comb mode (blue) enables soliton generation and the auxiliary mode (red) provides thermal compensation. In the dual-mode pumping scheme, the pump simultaneously couples to both modes by operating on the red-detuned side of a comb resonance and the blue-detuned side of an auxiliary resonance. (b) Standard microring configuration, in which pump coupling to both the comb mode (TE$_{00}$) and auxiliary mode (TE$_{10}$) is through the same bus-to-resonator coupler. (c) Proposed racetrack configuration, where the bus-to-resonator coupler is optimized for the comb mode, while engineered bending (adiabatic Euler bend with minimum radius $R_{\min}$) introduces controlled intermodal coupling to the auxiliary mode. The dashed box highlights the curved section where this coupling is induced.}
\end{figure*}

In long-cavity (low-repetition-rate) resonators, the higher threshold power leads to stronger thermo-optic instability, thereby requiring increased pump coupling to the auxiliary mode for effective thermal compensation. At the same time, the larger round-trip loss necessitates stronger pump coupling to the comb-generating mode to maintain near-critical coupling for efficient comb generation. Because the comb-generating and auxiliary modes typically exhibit different effective indices and loss characteristics, these simultaneous and increasingly stringent coupling requirements cannot be readily satisfied with a standard resonator configuration in which pump coupling to both modes relies on the same bus-to-resonator coupler. To overcome this limitation, we decouple the coupling pathways by using the bus coupler to address the comb-generating mode, while engineering the cavity geometry to independently control coupling to the auxiliary mode.

In this work, we introduce a passive, geometry-engineered approach to realize this concept. We employ a racetrack microresonator in which the curved sections are engineered to introduce controlled mode coupling between the comb-generating mode and an auxiliary mode, thereby enabling efficient pump coupling to the auxiliary mode for intrinsic thermal compensation during soliton formation. Using this fully passive approach in a high–quality-factor (Q>$10^7$) $\mathrm{Si_3N_4}$ microresonator, we demonstrate thermally accessible and deterministic single-soliton generation at a repetition rate of 33 GHz, with spectral coverage spanning the telecom C- and L-bands. This approach provides a straightforward and robust strategy for overcoming thermal instabilities in long-cavity microresonators and enables broadband, low-repetition-rate soliton microcomb generation.

\section{\label{sec:principle}Principle}

For soliton microcomb generation using the dual-mode pumping scheme \cite{Li2017StablyRegime, Weng2021DirectlyMicroresonator, Weng2022Dual-modeSolitons}, the microresonator is designed to support two spatial modes as shown in Fig.~\ref{fig:principle}(a), where the comb mode (blue) serves as the mode for soliton generation and the auxiliary mode (red) provides thermal compensation. During soliton formation, the pump (green arrow) operates on the red-detuned side of the comb-generating resonance \cite{Herr2014TemporalMicroresonators} and the blue-detuned side of an auxiliary resonance, enabling simultaneous coupling to both modes.

In conventional microring configurations (Fig.~\ref{fig:principle}(b)), pump coupling to both modes is mediated by a single bus-to-resonator coupler \cite{Weng2021DirectlyMicroresonator, Weng2022Dual-modeSolitons}, which limits independent optimization of the coupling strengths. As discussed above, low-repetition-rate operation requires increased pump coupling to both the comb-generating and auxiliary modes to ensure efficient comb generation and thermal compensation. Phase-matched coupler designs, such as symmetric directional couplers or asymmetric pulley couplers, can be used to efficiently couple the pump to a selected mode. However, because the comb-generating and auxiliary modes (e.g., TE$_{00}$ and TE$_{10}$) exhibit distinct effective indices, phase matching optimized for one mode generally leads to phase mismatch for the other, thereby limiting the ability to achieve sufficiently strong coupling to both modes simultaneously.

\begin{figure*}
\includegraphics[width=15.5cm]{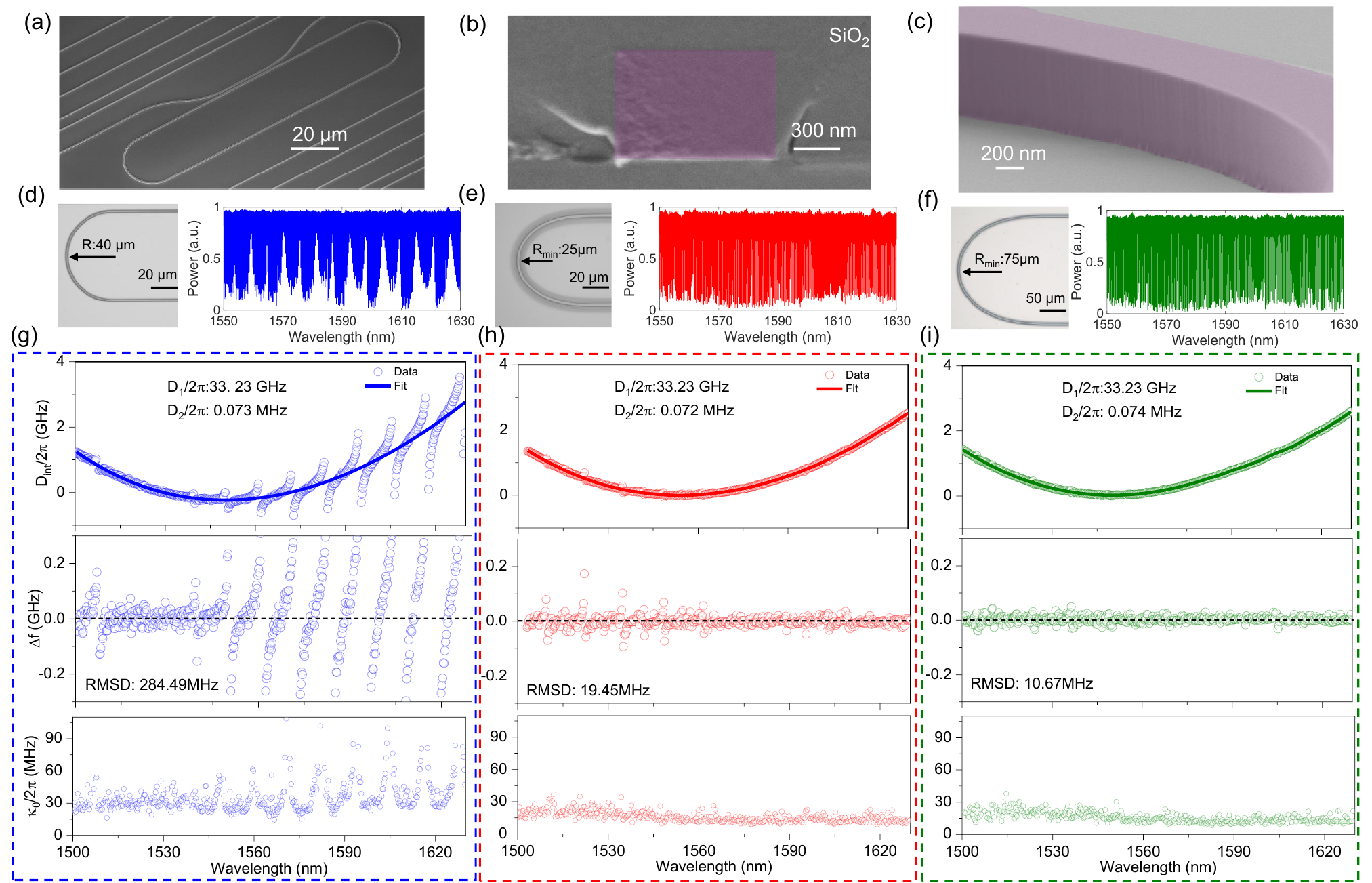}
\caption{\label{fig:design} \textbf{Mode coupling engineering in multimode racetrack resonators.} Scanning electron microscopy (SEM) image of the fabricated (a) racetrack microresonator, (b) its cross-section, and (c) sidewall. Proposed mode coupling engineering through (d) circular bends with 40 $\rm \mu m$ radius and Euler bends with minimum bending radius ($\rm R_{min}$) of (e) 25 $\rm \mu m$ and (f) 75 $\rm \mu m$ at curved sections. Left: optical microscope. Right: Transmission spectrum. (g, h, i) Corresponding linear characterization. First row: measured and fitted integrated dispersion ($\rm D_{int}/2\pi$). Second row: Resonance frequency deviation ($\rm \Delta f$) from the fitted curve, which reflects the mode coupling strength induced to the TE$_{00}$ mode. Third row: Intrinsic linewidth ($\rm \kappa_0/2\pi$) distribution across a broad wavelength range.}
\end{figure*}

To address this limitation, we introduce a configuration that decouples the coupling pathways for the two modes. While intermodal coupling is typically suppressed to prevent avoided-mode crossings\cite{Herr2014ModeMicroresonators, Kordts2016HigherFormation, Ji2021ExploitingCombs, Kim2021SuppressionMicroresonators, Ye2021IntegratedMicrocombs, Ji2022CompactCircuits, Ye2024MultimodePhotonics}, here we instead deliberately exploit controlled intermodal coupling to enable pump power transfer for thermal stabilization. The proposed racetrack configuration (Fig.~\ref{fig:principle}(c)) separates these functions: the bus coupler is optimized for the comb-generating mode, while the curved sections independently control coupling to the auxiliary mode. The dashed box highlights the engineered region where intermodal coupling is introduced.

Specifically, the curved waveguide employs an Euler bend with a tailored minimum bending radius ($R_{\min}$). Mode coupling arises when the transition between the straight and curved waveguides becomes non-adiabatic, which occurs when the modal evolution cannot follow the local waveguide geometry due to increased mismatch between the corresponding mode profiles. In conventional designs, a large $R_{\min}$ ensures a gradual transition and preserves adiabatic mode evolution, thereby minimizing intermodal interaction \cite{Ye2021IntegratedMicrocombs, Ji2022CompactCircuits}. Here, in contrast, $R_{\min}$ is intentionally reduced to enhance modal mismatch and shorten the effective transition length, driving the system into a controlled non-adiabatic regime that induces coupling from the comb-generating mode to the auxiliary mode.

A key design principle is that $R_{\min}$ is selected to be sufficiently small to enable intermodal coupling, yet sufficiently large to suppress bending-induced radiation loss and maintain the high-Q characteristics of the microresonator. This controlled interaction establishes a power exchange channel from the comb-generating mode to the auxiliary mode, providing an additional degree of freedom for thermal stabilization in low-repetition-rate soliton microcomb generation.

\section{\label{sec:results} Results}

\subsection{Device design and linear characterization}

To implement the proposed decoupled coupling scheme, we employ a racetrack microresonator consisting of straight and curved waveguide sections. The bus-to-resonator coupler is designed with identical waveguide dimensions to preserve high coupling ideality, and coupling to the comb-generating mode is achieved over a defined interaction length. The curved sections are engineered using an Euler-bend design to independently control coupling to the auxiliary mode. We designed three $\mathrm{Si_3N_4}$ racetrack microresonators with a multimode waveguide cross section (765~nm~$\times$~2~$\mathrm{\mu m}$) to intentionally introduce an auxiliary mode for mitigating thermal effects during soliton microcomb generation. To evaluate the proposed mode-coupling engineering approach, we fabricated both circular- and Euler-bend racetrack microresonators on the $\mathrm{Si_3N_4}$ platform using a subtractive process \cite{Ye2019High-QOptics, Ji2017, Zheng2025SiliconLasers}. A 765-nm-thick $\mathrm{Si_3N_4}$ layer was deposited on a silicon substrate with a thermally grown silicon dioxide buffer via low-pressure chemical vapor deposition. Device patterns were defined using optimized electron-beam lithography \cite{Zheng2018} and transferred into the $\mathrm{Si_3N_4}$ film via inductively coupled plasma reactive ion etching. The structures were then clad with silicon dioxide. Figures \ref{fig:design}(a, b, c) show scanning electron microscopy (SEM) images of the fabricated racetrack microresonator, its cross section, and its sidewall, respectively.

The left panels of Figs.~\ref{fig:design}(d, e, f) show that the circular-bend resonator features a fixed bending radius of 40~$\mathrm{\mu m}$, while the Euler-bend resonators have $\rm R_{min}$ of 25~$\mathrm{ \mu m}$ and 75~$\mathrm{\mu m}$. All devices have a cavity length of 4.36~mm, corresponding to an FSR of 33~GHz. The right panels of Figs.~\ref{fig:design}(d, e, f) present the measured transmission spectra of the three designs. The circular-bend resonator exhibits pronounced spectral modulation, while the Euler-bend designs show much smoother transmission, indicating weaker intermodal coupling. 

This difference is further confirmed in the first row of Figs.~\ref{fig:design}(g, h, i), which display the measured and fitted integrated dispersion $\rm \mathrm{D_{int}} = \sum \mu^n/(n!)D_n$, where $\mathrm{D_n}$ denotes the $n^{\mathrm{th}}$-order dispersion coefficient. The similar second-order dispersion ($\mathrm{D_2/2\pi}$) values among all devices confirm consistent waveguide dimensions. However, it is difficult to distinguish between the two Euler-bend designs based on $\rm D_{int}$ alone. Therefore, the second row of Figs.~\ref{fig:design}(g, h, i) plots the resonance-frequency deviation from the fitted curve, highlighting the AMX-induced mode shifts of the $\rm TE_{00}$ mode and the corresponding intermodal coupling strength.

The circular-bend resonator exhibits the strongest deviation, with a root-mean-square deviation (RMS $\rm \Delta f$) of 284.49~MHz, defined as $\sqrt{\sum (\Delta f)^2/n}$, where $n$ is the total number of measured modes. The periodicity of the spectral modulation indicates coupling between the TE$_{00}$ (comb-generating) and TE$_{10}$ (auxiliary) modes, arising from modal mismatch between straight and circular-bend sections. In contrast, the Euler-bend resonators show reduced coupling strength, which can be adjusted by $\rm R_{min}$. The smaller $\rm R_{min}$ exhibits weak modulation (RMS $\rm \Delta f$ of 19.45~MHz), while the larger $\rm R_{min}$ shows almost complete suppression (RMS $\rm \Delta f$ of 10.67~MHz). The corresponding maximum mode deviations ($\Delta f$) are 200~MHz and 40~MHz, respectively, at the AMX wavelengths, from which the coupling rate between the comb-generating and auxiliary modes can be estimated as $(2\pi \Delta f / \mathrm{FSR})^2$ \cite{helgason2021dissipative}. These results confirm that mode coupling in the Euler-bend design can be engineered while remaining substantially weaker than that induced by conventional circular bends.

Finally, the third row of Figs.~\ref{fig:design}(g, h, i) compare the intrinsic linewidth $\rm \kappa_0/2\pi$ distributions across a broad wavelength range. In the circular bending design, strong mode coupling around the AMXs leads to severe Q-factor (large $\rm \kappa_0/2\pi$) degradation. In contrast, the Euler-bend designs enable controlled intermodal coupling while preserving high Q performance. Notably, the entire device footprint can be as small as 0.15~$\mathrm{mm^2}$.

\begin{figure}[t]
\centering
\includegraphics[width=7.5cm]{fig_3.pdf}% Here is how to import EPS art
\caption{\label{fig:Q} \textbf{Linewidth and frequency deviation measurement.} (a) Resonance-linewidth characterization of racetrack microresonators with different bend designs. (b) Root-mean-square frequency deviation ($\rm \Delta f$) and intrinsic-Q measurements in Euler bends with different minimum bending radii ($\rm R_{min}$).}
\end{figure}

To verify that high $Q$ performance is preserved even in Euler bends with a small minimum bending radius, Fig.~\ref{fig:Q}(a) presents histograms of the intrinsic linewidths for the three designs. The most probable linewidths for the circular bend and the Euler bends with minimum radii of 25~$\rm \mu m$ and 75~$\rm \mu m$ are 26~MHz, 12~MHz, and 11.5~MHz, respectively, corresponding to intrinsic $Q$ factors of $\rm 7.5 \times 10^6$, $\rm 16.25 \times 10^6$, and $\rm 16.95 \times 10^6$. These results confirm that the high-$Q$ performance is maintained even with the introduction of controlled intermodal coupling. Fig.~\ref {fig:Q}(b) shows that the mode coupling strength can be tuned with respect to the minimum bending radius in the Euler-bend design, with the frequency deviation varying from approximately 21~MHz to 10~MHz while high Q performance is maintained. 

\subsection{Low-repetition-rate soliton microcomb generation} 

\begin{figure}[b]
\centering
\includegraphics[width=8.5cm]{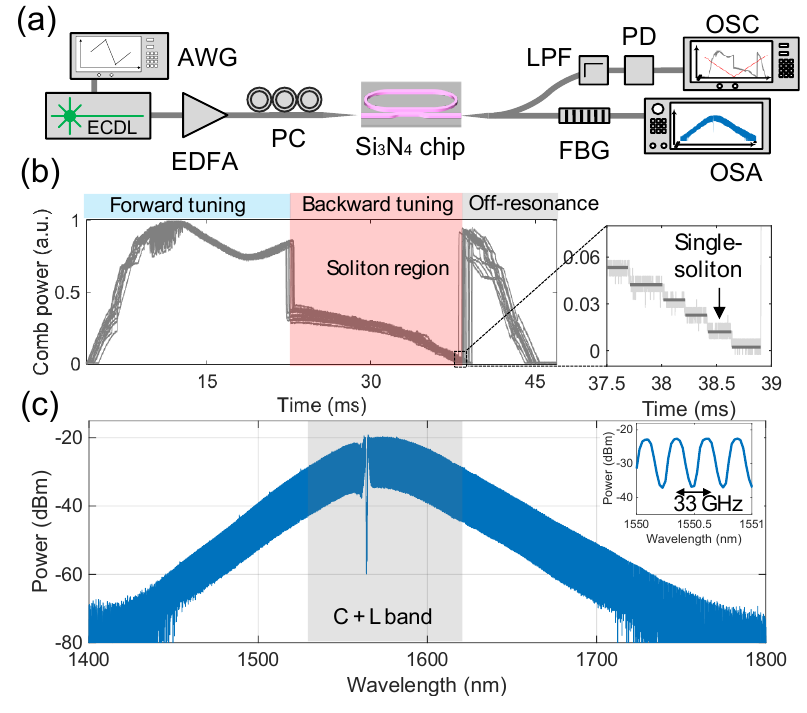}% Here is how to import EPS art
\caption{\label{fig:soliton} \textbf{Low-repetition-rate soliton microcomb generation.} (a) Schematic of the experimental setup. AWG: arbitrary waveform generator; ECDL: external-cavity diode laser; PC: polarization controller; EDFA: erbium-doped fiber amplifier; LPF: long-pass filter; PD: photodetector; OSC: oscilloscope; FBG: fiber Bragg grating; OSA: optical spectrum analyzer. (b) Twenty overlaid traces of the output comb power (gray) during forward (blue) and backward (red) tuning over the resonance, recorded with the same pump power and tuning speed. The right panel shows a magnified view of the backward tuning with successive switching of soliton states. (c) Optical spectrum of a broadband single-soliton comb with a repetition rate of 33~GHz, where the pump is filtered by an FBG.}
\end{figure}

As discussed in Section~\ref{sec:principle}, thermally accessible low-repetition-rate soliton microcomb generation benefits from microresonators with an appropriate mode-coupling strength. Therefore, the design in Fig.~\ref{fig:design}(e) is used for the experiment because it enables weak intermodal coupling while maintaining high-$Q$ performance. We first identify a $\mathrm{TE_{00}}$ resonance that is weakly coupled to a nearby higher-order $\mathrm{TE_{10}}$ mode located on the red-detuned side. This configuration provides thermal compensation when the pump laser is tuned across the resonance. The experimental setup is shown in Fig.~\ref{fig:soliton}(a). A continuous-wave (CW) pump, amplified by an erbium-doped fiber amplifier, is coupled to the $\rm TE_{00}$ mode of the $\mathrm{Si_3N_4}$ device. At the output, the comb power and spectrum are recorded by an oscilloscope and an optical spectrum analyzer, respectively. Figure~\ref{fig:soliton}(b) shows 20 overlaid traces of the transmitted comb power (gray) during forward (blue) and backward (red) tuning, recorded with the same on-chip pump power of 630~mW and a tuning speed of 100~GHz/s, indicating deterministic single-soliton generation. In practical operation, soliton states can be consistently accessed through manual tuning of the pump wavelength. The blue and red shaded regions denote laser operation on the blue- and red-detuned sides of the resonance, respectively, while the gray region marks the off-resonant regime. The right panel of Fig.~\ref{fig:soliton}(b) presents a magnified view of the backward-tuning trace, showing successive transitions from multiple- to single-soliton states.

The optical spectrum of the single-soliton microcomb is shown in Fig.~\ref{fig:soliton}(c), where the pump light is suppressed by a fiber Bragg grating. The spectrum spans approximately 300~nm, from 1450~nm to 1750~nm, with a repetition rate of 33~GHz (inset). The telecom C- and L-band regions are shaded in gray and include more than 300 comb lines with power levels exceeding --30~dBm. This performance is particularly attractive for ultra-long-haul optical transmission using the DWDM technique \cite{Swain2024SiliconCommunications}. For such systems, it is desirable to generate a densely spaced microcomb covering the C and L bands, with each comb line exceeding --30~dBm. Recent studies have demonstrated that, after 8000~km of transmission, the signal-to-noise ratio (SNR) difference between a commercial WDM laser bank and a microcomb source with --30~dBm input power is negligible \cite{Oxenlwe2025OpticalCommunications}. Achieving the target bandwidth and comb-line power typically requires operating substantially above the threshold power; however, the resulting increase in intracavity power exacerbates thermal instability and makes it difficult to stably access the soliton step. The proposed approach addresses this issue by employing engineered mode coupling to mitigate thermal effects, enabling deterministic soliton generation through simple manual tuning.

\section{\label{sec:conclusion}Conclusion}

In summary, we have demonstrated a thermally accessible and deterministic approach for generating low-repetition-rate soliton microcombs on the high-Q silicon nitride platform. By introducing engineered mode coupling through carefully designed Euler bends in racetrack microresonators, we achieved precise control over intermodal interaction while maintaining high-$Q$ performance. This controlled coupling effectively mitigates thermal instability during soliton formation under high pump power, enabling deterministic single-soliton generation through simple manual tuning. The resulting soliton microcomb exhibits a repetition rate of 33~GHz and a 300~nm optical bandwidth, covering the entire C and L bands with per-line power above --30~dBm, which is ideal for high-capacity coherent communication. The demonstrated results provide a simple and robust solution for realizing broadband, low-repetition-rate soliton microcombs. The proposed method can be extended to other platforms with a large thermo-optic coefficient, paving the way for applications in optical communications, microwave photonics, and precision metrology.

\textbf{Availability of data and materials}\\
The datasets generated and/or analyzed during this study are available from the corresponding authors upon reasonable request.

\textbf{Competing interests}\\
The authors declare no competing financial interests.

\textbf{Funding}\\
This work is supported by European Research Council under the EU's Horizon 2020 research and innovation programme (grant agreement no. 853522, REFOCUS), Horizon Europe research and innovation programme under the Marie Skłodowska-Curie grant agreement No. 101119968 (MicrocombSys), Independent Research Fund Denmark (ifGREEN 3164-00307A), Danish National Research Foundation (SPOC ref. DNRF123), and Innovationsfonden (Green-COM 2079-00040B).

%\nocite{*}%If you just want to include all the entries that have been cited

%\bibliographystyle{IEEEtran}
%\bibliographystyle{vancouver}
 \bibliography{references1}

% \bibliography{references, apssamp}% Produces the bibliography via BibTeX.$

\end{document}